\documentclass[preprint,showpacs,aps]{revtex4}
\usepackage{graphicx}
\usepackage{amssymb}

\def\RR{\hbox{{\rm I}\kern-.2em\hbox{\rm R}}}
\def\pRR{\hbox{{\tiny \rm I}\kern-.1em\hbox{{\tiny \rm R}}}}

\def\NN{\hbox{I\kern-.2em\hbox{N}}}

\begin{document}

\title{Asymptotic and numerical studies of the Becker-D\"oring model for
transient homogeneous nucleation}
\author{ L. L. Bonilla\cite{bonilla:email}}
\affiliation{Escuela Polit\'ecnica Superior, Universidad Carlos
III de Madrid, Avda.\ Universidad 30, E-28911 Legan{\'e}s, Spain}
\author{ A. Carpio\cite{carpio:email}}
\affiliation{Departamento de Matem\'atica Aplicada, Universidad
Complutense de Madrid, E-28040 Madrid, Spain}
\author{ Y. Farjoun\cite{farjoun:email}}
\affiliation{Department of Mathematics, University of California
at Berkeley, Berkeley, CA 94720, USA}
\author{ J. C. Neu\cite{neu:email}}
\affiliation{Department of Mathematics, University of California
at Berkeley, Berkeley, CA 94720, USA}
\date{\today}

\begin{abstract}
Transient homogeneous nucleation is studied in the limit of large critical sizes. Starting
from pure monomers, three eras of transient nucleation are characterized in the classic
Becker-D\"oring kinetic equations with the Turnbull-Fisher discrete diffusivity. After an
initial stage in which the number of monomers decreases, many clusters of small
size are produced and a continuous size distribution is created. During the second era,
nucleii are increasing steadily in size in such a way that their distribution
appears as a wave front advancing towards the critical size for steady
nucleation. The nucleation rate at critical size is negligible during this era.
After the wave front reaches critical size, it ignites the creation  of supercritical clusters
at a rate that increases monotonically until its steady value is reached. Analytical formulas
for the transient nucleation rate and the time lag are obtained that improve
classical ones and compare very well with direct numerical solutions. In addition, we propose
and solve numerically a modified Becker-D\"oring model having a discrete diffusivity
proportional to the area of a spherical cluster with $k$ monomers for small $k$ (as in
the Turnbull-Fisher case) and to the cluster radius for large $k$ (as in the case of diffusive
growth of clusters).\\
{\bf Key words:} Kinetics of first order phase transitions; homogeneous nucleation; 
Becker-D\"oring equations; singular perturbation; asymptotic theory.\\
{\bf AMS subject classifications.}  82C20; 82C26; 34E15.
\end{abstract}

\maketitle

\section{Introduction}
\label{sec-introduction}
In first order phase transitions, it is possible to achieve that the system is in a metastable
phase for values of the control parameter at which another phase is stable. Then nucleii of
the stable phase appear at random locations and they typically grow or decay depending on
whether their size surpasses a critical size \cite{ll10}. In homogeneous nucleation processes,
the probability density of finding a cluster of the stable phase with $k$ monomers does not
depend on the position. Homogeneous nucleation occurs in many examples of first order phase
transitions, such as condensation of liquid droplets from a supersaturated vapor, glass-to-crystal 
transformations \cite{kel83}, crystal nucleation in undercooled liquids \cite{kel91}, and 
in polymers \cite{cap03}, growth of spherical aggregates beyond the critical micelle 
concentration (CMC) \cite{isr91,neu02}, precipitation processes \cite{nie03} and the 
segregation by coarsening of binary alloys quenched into the miscibility gap \cite{LS,cap03}. 
In condensed systems, the time needed for the nucleation rate of supercritical clusters to reach 
a stationary value is large, which facilitates their theoretical and experimental study 
\cite{kel91}. Authoritative reviews of nucleation in condensed systems are due to Wu
\cite{wu96} for theory, analytical and numerical methods, and to Kelton \cite{kel91},
who is not always reliable in his assessment of theories, but included a wealth of very valuable
experimental data.

While nucleation is a random process, the cluster size distribution function satisfies
deterministic equations which are particular cases of coagulation-fragmentation equations
\cite{rezakhanlou}. Typically, we assume that a cluster can grow or decay by adding or 
shedding one monomer at a time. Then homogeneous nucleation is described by the 
Becker-D\"oring (BD) discrete kinetic equations \cite{kel91,pen83}. Suppose that 
nucleation occurs in a lattice in which there are many more binding sites, $M$, than particles, 
$N$, \cite{neu02,NBC}. We shall consider the thermodynamic limit, $N\to\infty$ with fixed 
particle density per site, $\rho\equiv N/M$. Let $p_{k}$ be the number of clusters with $k$ 
particles or, in short, $k$ clusters, and let $\rho_k\equiv p_k/M$ be the density of $k$ 
clusters. Note that the number densities per site, $\rho$ and $\rho_{k}$, are both 
dimensionless. Number densities per unit volume are obtained dividing $\rho$ and 
$\rho_{k}$ by the molecular volume, $v=V/M$. The BD equations (BDE) are \cite{neu02}
\begin{eqnarray}
\dot{\rho}_{k} = j_{k-1}- j_k\equiv - D_-\, j_k,
\quad k\geq 2,    \label{e2}\\
j_{k} = d_{k}\,\left\{ e^{{D_+\varepsilon_{k}\over k_B T}}\, \rho_1
\rho_k -  \rho_{k+1} \right\}.  \label{e3}
\end{eqnarray}
In (\ref{e3}), $\varepsilon_{k}$ is the binding energy of a $k$ cluster, required to
separate it into its monomer components. For spherical aggregates,
\begin{eqnarray}
\varepsilon_{k} = \left( (k-1) \alpha  - {3\over 2}\sigma
(k^{{2\over 3}} -1)\right)\, k_B T.  \label{e5}
\end{eqnarray}
This formula holds for $k\gg 1$, but we shall use it for all $k\geq 1$. $\alpha k_{B}T$
is the monomer-monomer bonding energy \cite{isr91} which, in the case of precipitation
of crystals from a solution or segregation by coarsening of binary alloys, may depend on the
particle density $\rho$ (volume fraction) through some empirical formulas \cite{pen83}.
In Eq.\ (\ref{e5}), $\sigma=2\gamma_{s} (4\pi v^2/3)^{{1\over 3}}/(k_B T)$,
where $\gamma_{s}$ and $v=V/M$ are the interfacial free energy per unit area (surface
tension) and the molecular volume, respectively. Note that $\alpha$ and $\sigma$ are
both dimensionless. The correction $3\sigma k_{B}T/2$ in (\ref{e5}) ensures that
$\varepsilon_{1}=0$, and it improves the agreement between the nucleation rate obtained
from the BDE and experiments \cite{wu96}. More precise atomic models were proposed
by Penrose et al \cite{pen83}.

The monomer density $\rho_{1}$ can be obtained from the conservation identity
\begin{eqnarray}
\sum_{k=1}^\infty k \rho_k = \rho, \label{e1}
\end{eqnarray}
in which the total particle density $\rho$ is constant. In (\ref{e2}), $\dot{\rho}_{k}
= d\rho_{k}/dt$ and $D_{\pm} u_k\equiv \pm [u_{k\pm 1}- u_{k}]$ are finite differences.
The time $t$, the discrete diffusivity $d_{k}$ and the flux $j_{k}$ are nondimensional. $t$
and $d_{k}$ are related to the dimensional time $t^*$ and decay coefficient $d^*_{k}$ as
follows \cite{NBC}
\begin{eqnarray}
t= \Omega t^*,\quad d_{k}= {d^*_k  \over\Omega}.  \label{e12}
\end{eqnarray}
Here the factor $\Omega$ has units of frequency, it depends on the particular model we
choose for $d_{k}$, and is determined in Appendix \ref{app:TF} for the Turnbull-Fisher
(TF) kinetics (which assumes that a monomer has to overcome an activation energy barrier for
its transfer across the interface of a cluster). The TF $d_{k}$ is
\begin{eqnarray}
d_{k}=  k^{2/3}\, e^{D_{+}g_{k}/2},\quad \Omega=\frac{12 D_{0}e^{-Q/(RT)}
}{v^{2/3}} .  \label{e18}
\end{eqnarray}
Here $D=D_{0} e^{-Q/(RT)}$ is the diffusion coefficient in the liquid, $Q$ is the
activation energy for diffusion, $R=k_{B}N_{A}$ is the gas constant and $v$ is the
molecular volume. In the classical theory, $d_{k}$ is proportional to the surface area of a $k$
cluster. In other models, $d_{k}$ is selected so as to yield the known expression for the
adiabatic growth of a nucleus of critical size by diffusion \cite{nie03} or by heat transfer
\cite{NBC}. The discrete diffusivity of these later models is proportional to the cluster
radius, thereby to $k^{1/3}$.

The flux $j_k$ in size space is the net rate of creation of a $k+1$ cluster from a $k$ cluster,
given by the mass action law. Notice that we have selected the kinetic coefficient for
monomer aggregation to $d_{k}$, the coefficient for decay of a $(k+1)$ cluster, so that
\begin{eqnarray}
\tilde{\rho}_k = \rho_{1}^k\, \exp \left({\varepsilon_{k}\over k_{B}T}\right).
\label{e4}
\end{eqnarray}
is the equilibrium size distribution solving $j_{k}=0$. This is the detailed balance assumption
whose validity is discussed in Wu's review \cite{wu96}.

Eqs.\ (\ref{e2}), (\ref{e3}), (\ref{e5}), (\ref{e1}) and (\ref{e18}) form a closed
system of equations that we can solve for an appropriate initial condition. If initially only
monomers are present, we have $\rho_{1}(0) = \rho$, and $\rho_{k}(0)=0$ for $k\geq 2$.
The stationary solutions of the BDE and the phenomenon of phase segregation \cite{BCP86}
will be described below in Section \ref{sec:model}. In Section \ref{sec:asymptotics}, we 
shall solve numerically the BDE for parameter corresponding to the glass-crystal transition in 
disilicate glasses \cite{kel91} and explain the results by means of an asymptotic theory 
valid for large critical size. A more detailed version of the material in Sections 
\ref{sec:model} and \ref{sec:asymptotics} of the present paper can be found in 
Ref.~\cite{NBC} which, in addition, contains a model for the kinetic constant $d_{k}$ 
based on thermal diffusion, not on activation processes as the TF model. The asymptotic 
theory for the discrete BDE is essentially the same for both models of the constant $d_{k}$. 
Most previous asymptotic theories correspond to the continuum limit of the BDE which is
a parabolic partial differential equation known as the Zeldovich-Frenkel equation (ZFE)
\cite{wu96}, and thus differ from and are less precise than ours \cite{NBC}. 

A somewhat puzzling point in nucleation theory is that the growth of a $k$ cluster is 
proportional to its surface area (therefore to $k^{2/3}$) if we assume that capture or emission 
of a monomer is an activated process as in the TF model. However, clusters are supposed to 
grow by diffusion (and thus at a rate proportional to $k^{1/3}$) after nucleation has stopped 
and coarsening begins \cite{pen97}. In Section \ref{sec:heat} of the present paper, we 
present a new modified BD model such that the growth of a cluster is proportional to 
$k^{2/3}$ for small size $k$ (as in the TF model) and to $k^{1/3}$ (as for the diffusive 
growth of precipitates) if $k$ is large. We discuss briefly the results obtained with these two 
models and their relation to other studies in the literature.

\section{BDE and their stationary solutions}
\label{sec:model}
\subsection{Equilibrium size distribution}
The equilibrium distribution (\ref{e4}) satisfies $j_k =0$ and it can be written as
\begin{eqnarray}
\tilde{\rho}_k = \rho_{1}\, e^{-g_{k}}, \label{e6}
\end{eqnarray}
where $g_{k}$ is the {\em activation energy} given by
\begin{eqnarray}
g_{k}= \sigma_{k} - (k-1)\, \varphi, \quad
\sigma_{k} =  {3\over 2}\sigma\, (k^{{2\over 3}}-1),\quad (k\geq 1),\quad
\varphi =\ln\left( e^\alpha \rho_{1}\right) .  \label{e8}
\end{eqnarray}
Here $\sigma_{1}= 0=g_{1}$. Assuming $k\gg 1$, $g_{k}$ achieves its global
maximum $g_{m}= \sigma k_{c}^{2/3}/2 + \sigma k_{c}^{-1/3}-3\sigma/2$ at the
critical size
\begin{eqnarray}
k = k_{c} \equiv \left({\sigma\over\varphi} \right)^3.  \label{e15}
\end{eqnarray}

Rewriting the flux (\ref{e3}) in the BDEs in terms of the activation energy, we obtain
\begin{eqnarray}
j_{k} &=& d_{k}\, \left\{ \left(e^{- D_+ g_{k}} - 1\right)\,
\rho_k - D_+ \rho_{k} \right\}.   \label{e13}
\end{eqnarray}
Eq.\ (\ref{e2}) is a spatially discrete Smoluchowski equation
with diffusion coefficient $d_{k}$ and drift velocity
\begin{eqnarray}
v_{k} = d_{k}\,\left(e^{- D_+ g_{k}} - 1\right).   \label{e14}
\end{eqnarray}
Notice that $v_{k}<0$ for $k<k_{c}$ (subcritical clusters shrink) and $v_{k}>0$ for
$k>k_{c}$ (supercritical clusters grow).

For the equilibrium densities (\ref{e6}), the conservation identity (\ref{e1}) becomes
\begin{eqnarray}
e^\alpha \rho = \sum_{k=1}^{\infty} k \left( e^{\alpha}\rho_{1}\right)^k\,
e^{-\sigma_{k}} = \sum_{k=1}^{\infty} k\, e^{k\varphi-\sigma_{k}}. \label{e10}
\end{eqnarray}
This series converges for $e^{\alpha}\rho_{1}=e^\varphi \leq 1$ ($\varphi\leq 0$),
and diverges for $e^{\alpha}\rho_{1} > 1$ ($\varphi>0$). At the critical micelle
concentration (CMC), $\rho_{1} = e^{-\alpha}$ ($\varphi=0$), we obtain the critical
density above which equilibrium is no longer possible,
\begin{eqnarray}
e^\alpha \rho_{c} = 1+ \sum_{k=2}^{\infty} k \, e^{-\sigma_{k}}. \label{e11}
\end{eqnarray}
For $\rho>\rho_{c}$, the BD kinetic equations predict phase segregation, i.e., indefinite
growth of ever larger clusters, and there remains a residual monomer concentration whose 
density $\rho_{1} e^\alpha\to 1$ as $t\to\infty$ \cite{BCP86,pen97,NBC}.

\subsection{The controlling parameters}
The simplest nucleation problem consists of solving the BD equations (\ref{e1}),
(\ref{e2}) and (\ref{e13}), with dimensionless activation energy $g_{k}= \sigma_{k}
- (k-1) \varphi$, discrete diffusivity $d_{k}$ (to be chosen later) and initial conditions
\begin{eqnarray}
\rho_1(0)=\rho,\,\, \rho_2(0) =\rho_3(0)=\ldots =0 . \label{e26}
\end{eqnarray}
The only parameters left in this initial value problem are $\rho$ and $\sigma$. $\rho$
controls the long-time behavior of the BDE: If $\rho\leq \rho_{c}$ given by (\ref{e11}),
$\rho_{k}(t)$ approach their equilibrium values  (\ref{e6}), with monomer density
$\rho_{1}$ that solves Equation (\ref{e1}). If $\rho> \rho_{c}$, cluster sizes grow
indefinitely whereas their density becomes small.

\begin{figure}
\begin{center}
\includegraphics[width=8cm]{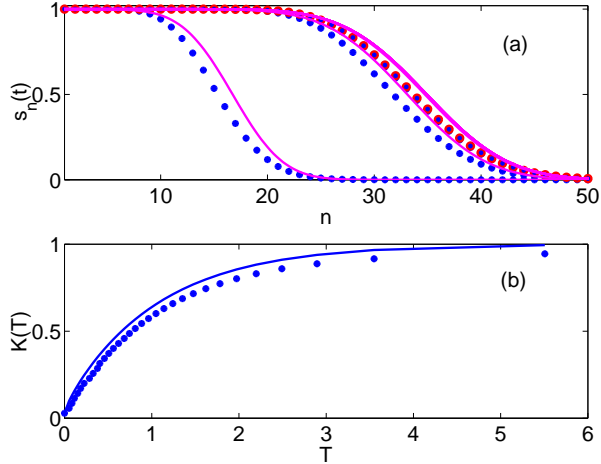}
\vspace{0.5 cm}
\caption{(a) Comparison of $s_{n}(t)$ evaluated (at different times) from the numerical
solution of the discrete equations (\ref{e31}) to the asymptotic result (\ref{a23}) (solid line).
(b) $K(T)$ calculated from Eq.\ (\ref{a9}) with $K(0)= \epsilon^3$ (solid line) is
compared to the numerically obtained position of the wave front. Data correspond to
disilicate glass at 820 K. All variables are written in dimensionless units.}
\label{fig2}
\end{center}
\end{figure}

Let us identify the controlling parameters $\rho$ and $\sigma$ in a physical system
undergoing homogeneous nucleation. A good experimental example for which abundant data
exist is the transformation of lithium disilicate glasses to crystals (devitrification) 
\cite{kel91}. In disilicate, the free energy per molecule of the crystal phase in the 
activation energy (\ref{e8}) is proportional to the undercooling
\begin{eqnarray}
\tilde{\varphi} = {\Delta S_{f} (T_{m}-T)\over N_{A}k_{B}T},  \label{g1}
\end{eqnarray}
where $T_{m}$ is the melting temperature, $\Delta S_{f}$ is the molar entropy of fusion
and $N_{A}$ is Avogadro's number; see the parameter values in Table I \cite{kel83}. The 
dimensionless density $\rho=e^{\varphi(0)-\alpha}$ can be extracted from Eq.\ 
(\ref{g1}) as explained in Section \ref{sec:asymptotics}. In energy units, the activation 
free energy is $k_{B}T g_{k} =\gamma_{s} 4\pi a^{2} - k_{B}T \varphi k$, where $a=
[3v/(4\pi)]^{1/3} k^{1/3}$ is the radius of a spherical $k$ cluster. Thus
\begin{eqnarray}
k_{B}T \,\left(g_{k} -\varphi + {3\sigma\over 2}\right) =
\gamma_{s} (4\pi)^{1/3} (3v)^{2/3} k^{2/3} - \Delta S_{f}(T_{m}-T) k/N_{A}.
\label{g2}
\end{eqnarray}
Comparing (\ref{g2}) with (\ref{e8}) yields $\sigma = (32\pi v^2/3)^{1/3}\gamma_{s}
/(k_{B}T)$, and the critical size
\begin{eqnarray}
k_{c}^{1/3}= \left({32 \pi v^2\over 3}\right)^{1/3} {\gamma_{s} N_{A}\over
\Delta S_{f}\, (T_{m}-T)} = \frac{\sigma}{\tilde{\varphi}}. \label{g3}
\end{eqnarray}
The other parameters in Table I will be used later to model the discrete
diffusivity in the BDE. We observe that the critical size increases with temperature: $k_{c}=
18$ at 703 K and $k_{c}= 34$ at 820 K. For other materials, such as undercooled liquid
metals, critical sizes can be rather large: liquid iron at maximum undercooling has $k_{c}=
494$, whereas $k_{c}= 2253$ for liquid rutenium at maximum undercooling \cite{kel91}.
\begin{table}[h]
\begin{center}
\begin{tabular}{*{3}{|c}|}    
\hline
Parameter & Symbol & Value\\
\hline
Melting temperature & $T_{m}$ & 1300 K\\
\hline
Entropy of fusion & $\Delta S_{f}$ & 40 J mol$^{-1}$ K$^{-1}$\\
\hline
Surface tension & $\gamma_{s}$  & 0.15 J/m$^2$\\
\hline
Preexponential diffusivity & $D_{0}$ & 2 $\times 10^9$ m$^2$ s$^{-1}$\\
\hline
Activation energy for diffusion & $Q$ & 440 kJ/mol\\
\hline
Molecular volume & $v$ & $10^{-28}$ m$^3$ \\
\hline
TF time scale (703K) & $\Omega^{-1}$  & 1.226 hours \\ 
 \hline
Critical size (703K) & $k_{c}$ & 18 \\
\hline
Undercooling (703K) & $\tilde{\varphi}$ & 4.087\\
\hline
Dimensionless surface tension (703K) & $\sigma$ & 10.74\\
\hline
Dimensionless free energy barrier (703K) & $g_{m}={\sigma\over 2}
k_{c}^{2/3} -{3\sigma\over 2} + \tilde{\varphi}$ & 25.177\\    \hline
 \end{tabular}
\label{tab1}
\end{center}
\caption{Data for lithium disilicate glass}
\end{table}

\subsection{Equivalent Becker-D\"oring system}
As they stand, the BDE are rather stiff and hard to solve numerically \cite{NBC}. This 
motivates the following change of variable
\begin{eqnarray}
\rho_{k} = \rho_{1} e^{-g_{k}} s_{k} = e^{-\alpha} e^{k\varphi-\sigma_{k}} s_{k},
\label{e28}
\end{eqnarray}
according to (\ref{e8}). Note that $s_{k}=1$ in equilibrium. Since $g_{1}=0$, this
equation implies
\begin{eqnarray}
s_{1}\equiv 1,    \label{e29}
\end{eqnarray}
for all $t$. For the initial condition (\ref{e26}), $e^{\varphi(0)-\alpha}=\rho_{1}(0) =
\rho$, and the conservation identity (\ref{e1}) becomes
\begin{eqnarray}
e^{\varphi(0)} = e^{\varphi} + \sum_{k=2}^\infty\, k\,
 e^{k\varphi -\sigma_{k}} s_{k}, \label{e30}
\end{eqnarray}
in which we have used (\ref{e28}). In terms of the $s_{k}$, the flux can be written as
\begin{eqnarray}
e^\alpha j_{k}= d_{k} \exp[(k+1)\varphi - \sigma_{k+1}]\, (s_{k}- s_{k+1}),
\label{flux}
\end{eqnarray}
and the BDE (\ref{e2}) and (\ref{e13}) become
\begin{eqnarray}
\dot{s_{k}} + u_{k} (s_{k+1} - s_{k}) = -k \dot{\varphi} s_{k}
+ d_{k-1}\, (s_{k-1} - 2s_{k} + s_{k+1}), \label{e31}
\end{eqnarray}
for $k\geq 2$. Here,
\begin{eqnarray}
u_{k} = d_{k-1} - d_{k} e^{\varphi - D_{+}\sigma_{k}}. \label{e32}
\end{eqnarray}
The term $u_{k}\, D_{+}s_{k}$ in Eq.\ (\ref{e31}) represents {\em discrete
advection}, with a drift velocity $u_{k} = - v_{k} +(d_{k-1} - d_{k})\sim - v_{k}$, which
is essentially minus the drift velocity in the original BDE for $k\gg 1$. Thus, the advection
 in Eq.\ (\ref{e31}) {\em climbs up} the activation energy barrier, from small values of
 $g_{k}$ to large ones.

 In summary, the transformed nucleation initial-boundary value problem consists of the
 balance equations (\ref{e31}), the particle conservation equation (\ref{e30}), the
 boundary condition (\ref{e29}), $s_{1}=1$, and initial conditions $s_{k}(0)=0$ for all
 $k\geq 2$. Its solution gives $\varphi(t)$ and $s_{k}(t)$ for all $k\geq 2$ and all $t>0$.

\subsection{Stationary solution}
The stationary solution of the BDE has a flux independent of cluster size, so that  $e^{\alpha}
j_{k}= d_{k} \exp[(k+1)\varphi - \sigma_{k+1}]\, (s_{k}- s_{k+1})= j$, from which
$(s_{k+1}- s_{k}) = - j \exp[\sigma_{k+1} - (k+1)\varphi]/d_{k}$, and therefore
 \begin{eqnarray}
s_{k}= 1-j \sum_{l=1}^{k-1} {\exp[\sigma_{l+1}-(l+1)\varphi]\over d_{l}},
\label{e33}
\end{eqnarray}
for $k\geq 2$. Since $s_{\infty}=0$, $j$ can be obtained from this expression in terms
of an infinite series
\begin{eqnarray}
j= {1\over\sum_{l=1}^{\infty} \exp[\sigma_{l+1}-(l+1)\varphi -\ln d_{l}]}.
\label{e34}
\end{eqnarray}
Substituting back this expression into (\ref{e33}), we obtain
 \begin{eqnarray}
s_{k}= 1- {\sum_{l=1}^{k-1} \exp[\sigma_{l+1}-(l+1)\varphi -\ln d_{l}]\over
\sum_{l=1}^{\infty} \exp[\sigma_{l+1}-(l+1)\varphi -\ln d_{l}]}.
\label{e35}
\end{eqnarray}
Then, $\rho_{k} = \rho_{1}\, e^{-g_{k}} s_{k}$.

\section{Asymptotic theory of nucleation}
\label{sec:asymptotics}
In this section, we shall interpret the numerical solutions shown in Figures \ref{fig2} and
\ref{fig4} by using singular perturbation methods. Starting from pure monomers, the
numerical solution of the BDE show that there are three well differentiated stages or eras of 
transient nucleation. After an initial stage in which the number of monomers $\rho_{1}$
decreases, many clusters of small size are produced and a continuous size distribution is 
created. During the second era, nucleii are increasing steadily in size in such a way that their 
continuum size distribution appears as a wave front advancing towards the critical size for 
stationary nucleation; see Fig.~\ref{fig2}. The nucleation rate at critical size is negligible 
during this second era. After the wave front reaches critical size, it ignites the creation of
supercritical clusters at a rate that increases monotonically until its steady value is reached; see
Fig.~\ref{fig4}. Our asymptotic theory of the BDE will be described using the TF discrete 
diffusivity (\ref{e18}) and compared to numerical solution of the BDE for the crystallization 
of disilicate glass at different undercoolings.

\subsection{Initial transient}
Initially, $\rho_1(0) = \rho$ and there are no multiparticle aggregates. There is an
initial transient stage during which dimers, trimers, etc.\ form at the expense of the
monomers. This initial stage is characterized by the decay of the chemical driving force
$\varphi=\alpha+\ln \rho_{1}$ to a quasi-stationary value $\tilde{\varphi}$, given
by Eq.\ (\ref{g1}) in the case of disilicate glass, and the emergence of a continuum size
distribution. Knowing this, {\em we choose the initial chemical driving force $\varphi(0)$ so
that the quasistationary value $\tilde{\varphi}$ given by Eq.\ (\ref{g1}) is attained
at the end of the initial stage.}

In materials such as disilicate glass at the temperatures we consider, the critical size is
relatively small. Then $\varphi(0)\approx\tilde{\varphi}$, and the initial stage is very
short. As the critical size increases (as in the case of undercooled liquid metals), $\varphi(0)$
may differ appreciably from $\tilde{\varphi}$, and the initial stage lasts longer. However,
even in such cases, the duration of the initial stage compared to the duration of the overall
transient to stationary nucleation is of order $k_{c}^{-2/3}\ll 1$ \cite{NBC}.

\subsection{Wave front advancing towards the cluster of critical size}
After the first era, clusters of increasing size are formed. For sufficiently small clusters,
the continuum size distribution approaches the equilibrium distribution with $\varphi=
\tilde{\varphi}$. This situation can be observed as an advancing wave front in the variable
$s_{k}(t)$, satisfying $s_{k}\sim 1$ (equilibrium) behind the front and $s_{k}\sim 0$
ahead of the front. This second era is described by Equations (\ref{e30}) to (\ref{e32})
with $\varphi=\tilde{\varphi}$ and $\dot{\varphi}=0$. The critical sizes (\ref{g3})
for disilicate glass are relatively small, between 10 and 50, but they are large for
undercooled liquid metals, generally between 100 and 1000. Hence we shall use $k_{c}^{-
1/3}$ as a small gauge parameter
\begin{eqnarray}
\epsilon = {\tilde{\varphi}\over\sigma}. \label{a2}
\end{eqnarray}
Our asymptotic analysis will be carried out in the limit $\epsilon\to 0$, and therefore
$k_{c}= \epsilon^{-3}\to\infty$. Then $d_{k}$, $u_{k}$ and $\sigma_{k}$ in Eqs.\
(\ref{e18}), (\ref{e31}) and (\ref{e32}) are smooth functions of $k>0$:
\begin{eqnarray}
d(k) = k^{2/3}\, e^{[D_{+}\sigma(k) - \tilde{\varphi}]/2},\quad
\sigma(k) = {3\over 2} \sigma\, (k^{2/3} -1),   \label{a3}\\
u(k) = d(k-1) - d(k)\, \exp[\tilde{\varphi} - \sigma(k+1) + \sigma(k)]. \label{a4}
\end{eqnarray}

\subsubsection{Position of the wave front}
In the numerical solutions shown in Fig.~\ref{fig2}(a), the graphs of  $s_{k}$ vs.\ $k$ at
fixed time have clear inflection points at some $k$, where $s_{k}\approx 1/2$. The
inflection point is taken as the {\em position} of the wave front. In the continuum model, the 
front position $k= k_{f}(t)$ is a smooth function which obeys $\dot{k}_{f} = u(k_{f})$.   
If we scale $k_{f}=K/\epsilon^3$, this equation becomes
\begin{eqnarray}
&& {dK\over dT} = U(K),   \quad T=\epsilon t,   \label{a9}\\
&& U(K) \equiv  \lim_{\epsilon\to 0} [\epsilon^2 u(\epsilon^{-3} K)] =
2 K^{2/3}\, \sinh\left( {\tilde{\varphi}\over 2} (K^{-1/3}-1) \right) ,\label{a6}
\end{eqnarray}
in the limit as $\epsilon\to 0$. Fig.~\ref{fig2}(b) compares the wave front 
position calculated by solving (\ref{a9}) with $K(0)=\epsilon^3$
to the numerical solution of (\ref{e31}). Note that the solution of (\ref{a9}) presents 
a time shift with respect to the numerical solution of the discrete model. This time
shift reflects the breakdown of the continuum limit as $K\to 0$, due
to discreteness, and also the transient in $\varphi(t)$ before it settles to 
$\tilde{\varphi}$. If the solution of Eq.\ (\ref{a9}) - (\ref{a6}) is forced 
to agree with the numerical $K(T)$ when the latter is, say, 0.1, the comparison fares much 
better.

\subsubsection{Shape of the wave front}
The leading edge of the wave front is a layer centered at $K(T)$ in which $s_k$ decreases
from 1 to 0 as $k$ increases through it. The continuum representation of $s_k$ in this layer is
\begin{eqnarray}
s_{k} = S(X,T;\epsilon), \quad X = \epsilon^{3/2} \left( k - {K\over \epsilon^3}
\right).    \label{a11}
\end{eqnarray}
Inserting (\ref{a11}) into (\ref{e31}), and then using (\ref{a9}), we obtain
\begin{eqnarray}
&& {\partial S\over\partial T} + U'(K) X {\partial S\over\partial X}
= D(K)\, {\partial^2 S\over\partial X^2} , \label{a16}\\
&& D(K)\equiv \lim_{\epsilon\to 0}\left[d(\epsilon^{-3}K)- {1\over 2}\,
u(\epsilon^{-3}K)\right] \epsilon^2  = K^{2/3} \cosh\left( {\tilde{\varphi}
\over 2}(K^{-1/3}-1) \right),  \label{a17}
\end{eqnarray}
in the limit as $\epsilon\to 0$. Had we carried out the same analysis for the ZFE, we would have found $D(K)\sim d(
\epsilon^{-3}K)\,\epsilon$. This would have resulted in a wider wave front and a longer
time to ignition than those described below.

\subsubsection{Flux and wave front width}
Besides determining the shape of the wave front near its location, Eq.\ (\ref{a16})
yields the behavior of the flux (creation rate of clusters larger than $k$) $j_{k}$ near $k=
k_{f}$. If we substitute (\ref{e18}) and (\ref{a11}) into (\ref{flux}), we obtain
\begin{eqnarray}
j_k \sim \epsilon^{-1/2}\, K^{2/3} \, e^{3\tilde{\varphi}/(2\epsilon)}
\exp\left[- {G(K)\over\epsilon^3} - {G'(K)X\over\epsilon^{3/2}}
- {G'(K)\over 2} - {G''(K)\over 2}\, X^2\right]\, {\partial S\over\partial X}.\,
\label{a18}
\end{eqnarray}
Here, $ G(K) \equiv \tilde{\varphi}\, \left({3\over 2} K^{2/3} -
K\right) $ is a scaled version of the activation energy (\ref{e8}).

Since $j_k$ is proportional to $\partial S/\partial X$, it is convenient to differentiate
(\ref{a16}) with respect to $X$ in order to obtain an equation for $J\equiv -
\partial S/\partial X$,
\begin{eqnarray}
{\partial J\over\partial T} + U'(K)\, {\partial (X\, J)\over\partial X} =
D(K)\, {\partial^2 J\over\partial X^2}. \label{a20}
\end{eqnarray}
Notice that $J$ is locally conserved, and the following integral conservation identity holds:
\begin{eqnarray}
1 = - [S]_{-\infty}^\infty = - \int_{-\infty}^\infty {\partial S\over\partial X}
dX = \int_{-\infty}^\infty J\, dX. \label{a21}
\end{eqnarray}
Eq.\ (\ref{a20}) has Gaussian solutions satisfying (\ref{a21}),
\begin{eqnarray}
J(X,T) = {1\over 2\sqrt{\pi A(T)}}\,\exp\left[ -{X^2\over 4\, A(T)}\right],
\label{a22}
\end{eqnarray}
which yields
\begin{eqnarray}
S(X,T) = {1\over 2}\,\mbox{erfc}\left[ {X\over 2\sqrt{A(T)}}\right] \label{a23}
\end{eqnarray}
for the wave front profile \cite{NBC}. Inserting Eq.\ (\ref{a22}) in Eq.\ (\ref{a20}), 
we find the following equation for $A(T)>0$:
\begin{eqnarray}
{dA\over dT} &-& 2\, U'(K)\, A = D(K). \label{a24}
\end{eqnarray}

After insertion of (\ref{a22}), the flux (\ref{a18}) becomes
\begin{eqnarray}
&&\frac{j_k}{j_{\infty}} \sim \sqrt{{3\over 2A\tilde{\varphi}}}\, K^{2/3}\,
\exp\left\{ {\tilde{\varphi}\over 2\epsilon^3}- {G(K)\over\epsilon^3} -
{G'(K)X\over\epsilon^{3/2}}-{G'(K)\over 2} -\left[{G''(K)\over 2}+ {1\over 4A}
\right]\, X^2\right\}, \quad\, \label{a25}\\
&& j_{\infty}= \sqrt{{\tilde{\varphi}\over 6\pi\epsilon}}\,
\exp\left(- {\tilde{\varphi}\over 2\epsilon^3} + {3\tilde{\varphi}\over 2
\epsilon}\right).    \label{a27}
\end{eqnarray}
Here $K=K(T)$ and $A=A(T)$ are found by solving the differential equations (\ref{a9})
and (\ref{a24}) with initial conditions $K(0)=\epsilon^3$ and $A(0)= 3\epsilon^4/(2
\tilde{\varphi})$, respectively. Eq.\ (\ref{a27}) is the classical Zeldovich quasi-steady 
nucleation rate of supercritical clusters, and it can be directly obtained from the stationary 
flux (\ref{e34}) in the limit as $\epsilon\to 0$.

\subsection{The nucleation rate of supercritical clusters}
Let us now study the transient creation rate, in which $j\equiv j_{k_{c}}$ increases from 0
to the steady Zeldovich value (\ref{a27}). As we have just seen, our theory predicts that
{\em the wave front profile is given by (\ref{a23}), where $K(T)$ and $A(T)$ are solutions
of Eqs.\ (\ref{a9}) and (\ref{a24}), respectively. The flux of clusters with sizes larger
than $k$ is then given by Eq.\ (\ref{a25})}. Setting $k=k_{c}=\epsilon^{-3}$ (critical
size) and $X=(1-K(T))/\epsilon^{3/2}$ in this equation, we obtain the nucleation rate
predicted by our theory, $j(t)$. Its integral over time yields the number of supercritical
clusters, $N_{c}(t)$. We shall consider now a different and more explicit approximation of
these results: We linearize the wave front equation (\ref{a9}) about the critical size $K=1$
and insert the solution into (\ref{a25}). 

\subsubsection{Linearization of the wave front speed about the critical size}
Let us fix $k=k_{c}=\epsilon^{-3}$ (critical size) in the definition (\ref{a11}) of $X$:
\begin{eqnarray} 
X = {1-K\over \epsilon^{3/2}} \equiv \kappa.    \label{a28} 
\end{eqnarray}
We now set $X=\kappa$ in (\ref{a25}) and perform the limit as $\epsilon\to 0$ with
$\kappa$ fixed. The result is
\begin{eqnarray} 
j \sim  j_{\infty} e^{- \tilde{\varphi}\kappa^2/6- \epsilon^{3/2}
\tilde{\varphi}\kappa/6} \sim  j_{\infty} e^{- \tilde{\varphi}\kappa^2/6},  
 \label{a29} 
\end{eqnarray}  
provided we use the limiting stationary value $(4A)^{-1} = - G''(1)/2$.

The transient turns on when $\kappa\equiv (1-K)/\epsilon^{3/2} = O(1)$. Since 
$U(1-\epsilon^{3/2}\kappa) \sim \epsilon^{3/2}\tilde{\varphi}\kappa/3$, 
the wave front equation (\ref{a9}) yields
\begin{eqnarray} 
{d\kappa\over dT} = - {\tilde{\varphi}\over 3}\,\kappa,     \label{a30} 
\end{eqnarray}
as $\epsilon\to 0$. The solution of this equation is 
\begin{eqnarray} 
\kappa &=& \kappa_{M}\, e^{- \tilde{\varphi} e^{\tilde{\varphi}}(T-T_{M})/3}
= \kappa_{M} e^{-(t-t_{M})/(2\tau)},     \label{a42}\\  
\tau^{-1} &=& {2\over 3} \tilde{\varphi} \epsilon. \label{a40}
\end{eqnarray}
We select $\kappa_{M}$ as the value at which the flux $j$ reaches its inflection point. Then 
we may consider that the wave front has ignited the nucleation of supercritical clusters. 
Straightforward use of Eqs.\ (\ref{a29}) and (\ref{a30}) shows that 
 \begin{eqnarray}
\kappa_{M} =\sqrt{{6\over\tilde{\varphi}}}.      \label{a41}
\end{eqnarray}
Moreover, $T_{M} = \epsilon t_{M}$ is the {\em time to ignition}, at which the wave 
front $K(T)$ reaches the value $K=1-\epsilon^{3/2}\kappa_{M}$. From (\ref{a9}), we
obtain $t_{M}$ and from (\ref{a29}) and (\ref{a42}), we obtain the flux \cite{NBC}
\begin{eqnarray}
&& j \sim  j_{\infty} e^{- (t-t_{M})/\tau}, \label{a32}\\
&& t_{M} = t_{\infty} + {3 \over 2 \tilde{\varphi}\epsilon}
\left\{ \ln\left({\tilde{\varphi}(1-\epsilon^3)^2\over 6 \epsilon^3}\right)
\right.\nonumber\\
&&Ê\quad + \left.
\int_{\epsilon^3}^{1-\epsilon^{3/2}\kappa_{M}} \left[{\tilde{\varphi}\over
3 K^{2/3}\sinh\left[{\tilde{\varphi}\over 2}(K^{-1/3} -1)\right] }
+ {2\over K-1}\right] dK\right\}. \label{a34}
\end{eqnarray}
Here $t_{\infty}$ is the duration of the initial stage. Integrating $j(t)$ over time, we find
the number of supercritical clusters as a function of time. In the limit as $t\to\infty$, this
number is $N_{c}(t)\sim j_{\infty}\,( t-\theta)$, where the time lag $\theta$ is
approximately given by $\theta= t_{M} + \tau\gamma+ \tau E_{1}(e^{t_{M}/\tau})$,
in which $\gamma = 0.577215\ldots$ is Euler's constant and $E_{1}(x)$ is an exponential
integral, see the derivation in Appendix B of Ref.~\cite{NBC}. The time lag $\theta$ can
be directly compared to experimental values \cite{kel91}.

\begin{figure}
\begin{center}
\includegraphics[width=9cm]{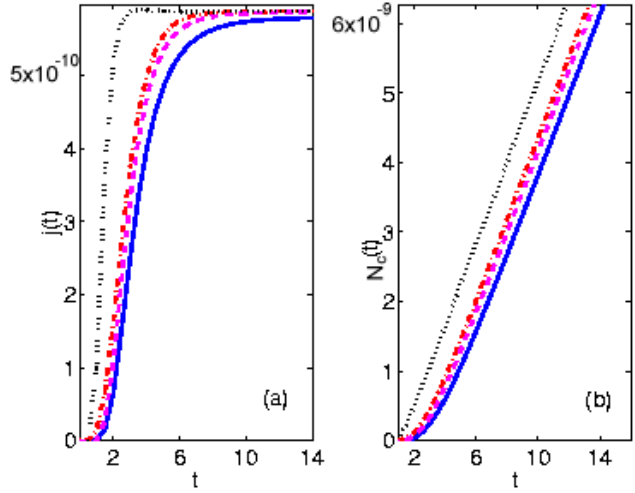}
\vspace{0.5 cm}
\caption{(a) Evolution of the dimensionless flux at critical size $j(t)$, and (b) number
of clusters surpassing critical size $N_{c}(t)$ as a function of dimensionless time
for disilicate glass at 703K, $k_{c}=18$. Solid lines correspond to numerical results, dashed
lines to the approximation given by Eq.\ (\ref{a25}), dot-dashed lines to the linearization
approximation (\ref{a32}) and dotted lines to linearizing the equations for $K(T)$ and
$A(T)$ as in Appendix C of Ref.~\cite{NBC}. }
\label{fig4}
\end{center}
\end{figure}

\subsubsection{Comparison between different approximations}
Fig.~\ref{fig4}(a) compares $j(t)$ calculated from the numerical solution of the BDE for
devitrification of disilicate glass at 703 K, from (\ref{a32}) and (\ref{a34}) with
$t_{\infty}=0$, and from Eq.\ (\ref{a25}) with $X=(1-K(T))/\epsilon^{3/2}$. We
find that the more precise expression, Eq.\ (\ref{a25}), captures better the width and
location of the transition region between $j=0$ and $j=j_{\infty}$, as compared with the
simple approximation given by Eqs.\ (\ref{a32}) and (\ref{a34}). Both approximations
present a small overshoot and yield a smaller time lag $\theta$ than that obtained from
the numerical solution of the BDE. The overshoot decreases as the critical size
increases. Another approximation consists of linearizing the equations for $K(T)$ and
$A(T)$ about the critical size $K=1$ as suggested in Ref. \cite{dem93}. This latter
approximation is the worst one. This is not surprising as such approximation provides the 
same result for both the discrete BDE and the continuum ZFE.

For disilicate glass at a lower temperature of 703 K, the critical size is smaller and
our approximations deviate more from the numerical solution of the BDE, as shown in
Figure \ref{fig4}(a). Integrating $j(T)$ over time, we find the number of supercritical
clusters as a function of time, $N_{c}(t)$, which is depicted in Figs. \ref{fig4}(b). 
The numerical solution of the BDE with the TF diffusivity yields
a time lag $\theta= 2.6$. This value is close to those provided by the linearization
approximation, $\theta=2.2$, and by Eq.\ (\ref{a25}), $\theta=2.3$. Thus these
analytical approximations to the numerical solution are reasonably good even for a
relatively small critical size. However, $\theta=2.6$ gives 3.2 hours according to
Table I, whereas the experimentally measured time lag is about 50 hours, cf.\ Fig.\ 5
of Ref.~\cite{kel91}. This discrepancy is due to having used the TF discrete
diffusivity, which yields an excessively small time unit, as shown in Table I.

\section{Modified model of nucleation and growth}
\label{sec:heat}
It is somewhat paradoxical that the TF discrete diffusivity is proportional to $k^{2/3}$
(cluster area), whereas growth stages of a cluster after nucleation are due to diffusive
accretion of monomers which yields a discrete diffusivity proportional to $k^{1/3}$ (cluster
radius). In this section, we propose a model that interpolates between these two mechanisms
and produces a discrete diffusivity proportional to $k^{2/3}$ for small cluster size and
proportional to $k^{1/3}$ for large cluster size. The idea is to consider that the discrete
diffusivity should be consistent with adiabatic growth of a large cluster at whose surface the
concentration of monomers is different from the monomer concentration at infinity. We then
modify the Becker-D\"oring equations to accomodate the resulting law for cluster growth 
\cite{yossi}.

\subsection{Modified Becker-D\"oring model}
To be precise, consider a spherical $k$-cluster which is growing adiabatically by diffusion of
monomers across its interface. If the diffusion coefficient is approximately constant, the
concentration solves Laplace's equation in spherical coordinates with densities $\rho_{a}$
at the sphere radius $r=a$, and $\rho_{\infty}$ at infinity:
\begin{eqnarray}
\rho(r)= \rho_{\infty} + \frac{\rho_{a}-\rho_{\infty}}{r}\, a. \label{ng1}
\end{eqnarray}
Suppose now that the crystal particles occupy all available sites inside the cluster, whereas
they occupy the volume fraction $\rho$ outside the cluster. Then the number of molecules
falling into the cluster surface per unit time is $(1-\rho)4\pi a^2 (da/dt)/v$, and this number
should equal the diffusive flux times the cluster surface area:
\begin{eqnarray}
\frac{4\pi a^2 D}{v}\, \frac{\partial\rho}{\partial r}(a)= (1-\rho)\, \frac{4\pi
a^2}{v}\,\frac{da}{dt^*}.   \label{ng2}
\end{eqnarray}
This implies
\begin{eqnarray}
\frac{dk}{dt^*}= \frac{(\rho_{\infty}-\rho_{a})\, 4\pi a D}{(1-\rho)\, v}=
 \frac{(\rho_{\infty}-\rho_{a})\, (4\pi )^{2/3} D (3k)^{1/3}}{(1-\rho)\, v^{2/3}}.
 \label{ng3}
\end{eqnarray}
Using now (\ref{tf16}) for large, near-critical cluster sizes,
\begin{eqnarray}
\frac{dk}{dt^*}\sim - \frac{12 D k^{2/3} D_{+}g_{k}}{v^{2/3}}\sim
- 12 D (k/v)^{2/3}\, (\sigma k^{-1/3}-\varphi_{a}),   \label{ng4}
\end{eqnarray}
in which $\varphi_{a}$ is the free energy per molecule at the cluster radius. Similarly,
$\rho_{\infty}-\rho_{a}\sim \tilde{\varphi}-\varphi_{a}$ if the free energy
per molecule is small, and Equations (\ref{ng3}) and (\ref{ng4}) yield
\begin{eqnarray}
\varphi_{a} - \sigma k^{-1/3} \sim \frac{(\pi/6)^{2/3}(\tilde{\varphi}-
\varphi_{a})}{(1-\rho)\, k^{1/3}}\Longrightarrow \varphi_{a} \sim \frac{
\tilde{\varphi}\, (1+\mu)}{\mu + \epsilon k^{1/3}}.   \label{ng5}
\end{eqnarray}
Here we have used (\ref{a2}) and the definition
\begin{eqnarray}
\mu = \frac{\epsilon\, (\pi/6)^{2/3}}{1- \rho}. \label{ng6}
\end{eqnarray}
Equations (\ref{a2}), (\ref{ng4}) and (\ref{ng6}) provide
\begin{eqnarray}
\frac{dk}{dt^*}\sim - \Omega k^{2/3}\frac{\tilde{\varphi} [(\epsilon k^{1/3})^{-1}
-1]}{1 + \epsilon k^{1/3}/\mu},   \label{ng7}
\end{eqnarray}
in which $\Omega$ is as defined in (\ref{e18}).

Equation (\ref{ng7}) describes the diffusive growth of a crystal nucleus in a glass phase
such that the concentration at the interface is different from that at infinity. We shall now
modify the BDE in such a way that the same equation is obtained in the limit of large
clusters near critical conditions. We shall replace the BD flux (\ref{e3}) by:
\begin{eqnarray}
j_{k} = \left( e^{ - D_{+} g_{k} + \Psi_{k}}\rho_{k} -\rho_{k+1}\right)\, d_{k},
\label{ng8}
\end{eqnarray}
which is equivalent to replacing $\rho_{1}e^{\Psi_{k}}$ instead of the monomer
density in (\ref{e3}). We shall now select $\Psi_{k}$ so that the new drift velocity
\begin{eqnarray}
v_{k} = \left( e^{ - D_{+} g_{k} + \Psi_{k}} -1\right)\, d_{k},    \label{ng9}
\end{eqnarray}
becomes
\begin{eqnarray}
v_{k} \sim - k^{2/3}\,\frac{ D_{+} g_{k}}{1 + \epsilon k^{1/3}/\mu},
\label{ng10}
\end{eqnarray}
in the limit of large, near critical cluster sizes. Writing time in dimensional units, Eq.\
(\ref{ng10}) is (\ref{ng7}) up to higher order terms. From (\ref{e18}) and (\ref{ng9}),
we obtain $v_{k}\sim - k^{2/3}\, (\Psi_{k} - D_{+}g_{k})$, which compared to
(\ref{ng10}) yields
\begin{eqnarray}
\Psi_{k} = \epsilon k^{1/3}\,\frac{ D_{+} g_{k}}{\mu + \epsilon k^{1/3}}.
\label{ng11}
\end{eqnarray}

The modified BDE are therefore (\ref{e2}) with the flux (\ref{ng8}) and (\ref{ng11})
together with the mass constraint (\ref{e1}).

\subsection{Equilibrium and stationary solution}
Our modified BDE do not have a Gibbsian equilibrium (\ref{e6}) nor satisfy detailed
balance. Instead, the equilibrium distribution for which $j_{k}=0$ is
\begin{eqnarray}
\tilde{\rho_{k}} = \exp\left( - g_{k}+ \sum_{j=1}^{k-1}\Psi_{j} \right).
\label{ng12}
\end{eqnarray}
As before, it is convenient to define a new distribution $s_{k}$ which is 1 at equilibrium:
\begin{eqnarray}
\rho_{k} = \exp\left( - \alpha +k\varphi - \sigma_{k}+ \sum_{j=1}^{k-1}\Psi_{j}
\right)\, s_{k}, \label{ng13}
\end{eqnarray}
for $k=2,3,\ldots$ and $s_{1}\equiv 1$. In terms of $s_{k}$, the flux (\ref{ng8})
becomes
\begin{eqnarray}
j_{k} = - d_{k} \exp\left(-\alpha + (k+1)\varphi - \sigma_{k+1} + \sum_{j=1}^k
\Psi_{j}\right)\, (s_{k+1}-s_{k}).    \label{ng14}
\end{eqnarray}
The stationary solution satisfies the equation $j_{k}= e^{-\alpha} j_{\infty}$, with
constant flux $j_{\infty}$. The result is
\begin{eqnarray}
s_{k} = 1- j_{\infty}\sum_{l=1}^{k-1} \exp\left(\sigma_{l+1}- (l+1) \varphi -
 \sum_{j=1}^l \Psi_{j}- \ln d_{l}\right).   \label{ng15}
\end{eqnarray}
Since $s_{k}\to 0$ as $k\to\infty$, we obtain the stationary flux
\begin{eqnarray}
j_{\infty}= {1\over\sum_{l=1}^{\infty} \exp[\sigma_{l+1}-(l+1)\varphi -
\sum_{j=1}^l \Psi_{j} - \ln d_{l}]}.  \label{ng16}
\end{eqnarray}

\subsection{Numerical results}
In terms of the $s_{k}$, the BDE are
\begin{eqnarray}
\dot{s_{k}} + u_{k} (s_{k+1} - s_{k}) = -\left( k \dot{\varphi} + \sum_{j=1}^{k-1}
\dot{\Psi}_{j}\right)\, s_{k} + d_{k-1}\, (s_{k-1} - 2s_{k} + s_{k+1}), \label{ng17}
\end{eqnarray}
for $k\geq 2$ and $s_{1}\equiv 1$. Here the new advection velocity is
\begin{eqnarray}
u_{k} = d_{k-1} - d_{k} e^{\varphi - D_{+}\sigma_{k}+\Psi_{k}}. \label{ng18}
\end{eqnarray}
We have to solve the BDE with initial condition $s_{k}(0)=0$ for $k\geq 2$. The mass
constraint (\ref{e30}) is now
\begin{eqnarray}
e^{\varphi(0)} = e^{\varphi} + \sum_{k=2}^\infty\, k\,
 e^{k\varphi -\sigma_{k}+\sum_{j=1}^{k-1}\Psi_{j}} s_{k}. \label{ng19}
\end{eqnarray}
In (\ref{ng11}), the parameter $\mu$ can be written as
\begin{eqnarray}
\mu = \frac{\epsilon\, (\pi/6)^{2/3}}{1- e^{\varphi(0)-\alpha}}, \label{ng20}
\end{eqnarray}
because $\rho=\exp[\varphi(0)-\alpha]$.

As indicated previously, with the initial condition of pure monomers $\varphi$ rapidly
evolves to a quasistationary value $\tilde{\varphi}$ and a continuum size distribution
emerges. We have to select the initial chemical driving force $\varphi(0)$ so that the
quasistationary value $\tilde{\varphi}$ given by Eq.\ (\ref{g1}) is attained at the end
of the initial stage. Notice that $\varphi(0)$ enters the definition (\ref{ng20}) and the
constraint (\ref{ng19}). Not knowing the monomer-monomer bonding energy $\alpha$, 
we find easier to fix $\mu$ at different values and find $\varphi(0)$ in (\ref{ng19}) so that 
the quasistationary value $\tilde{\varphi}$ of Eq.\ (\ref{g1}) is attained at the end
of the initial stage. The limit as $\mu\to \infty$ gives us back the original BDE with the
TF discrete diffusivity.

\begin{figure}
\begin{center}
\includegraphics[width=8cm]{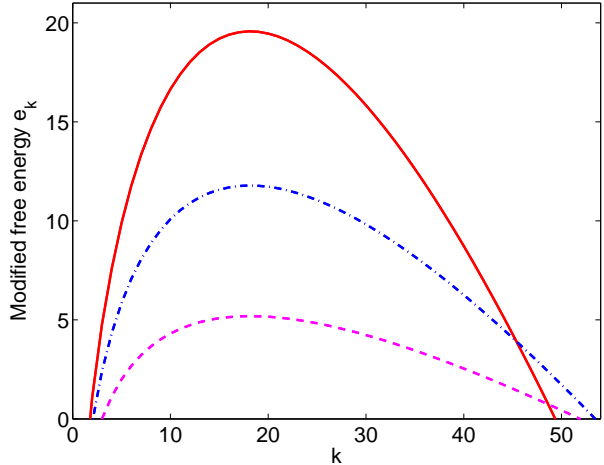}
\vspace{0.5 cm}
\caption{ Modified free energy $e_{k}=g_{k}- \sum_{j=1}^{k-1}\Psi_{j}-\alpha$
as a function of the size $k$ for $\mu=10$ (solid line), $\mu=1$ (dot-dashed line) and $\mu=
1/3$ (dashed line). }
\label{figm1}
\end{center}
\end{figure}

\begin{figure}
\begin{center}
\includegraphics[width=8cm]{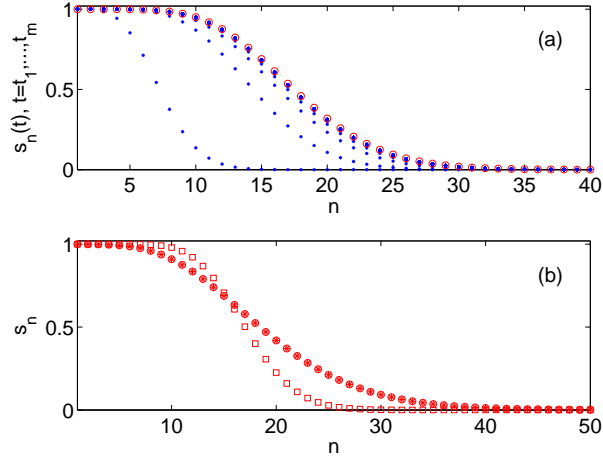}
\vspace{0.5 cm}
\caption{(a) Profiles of $s_{k}$ for $\mu=1$ at different times. The profile of the stationary
size distribution function is indicated by circles. (b) Stationary profiles of the size
distribution function for $\mu=10$ (squares), $\mu=1$ (circles) and $\mu=1/3$ (dots).}
\label{figm2}
\end{center}
\end{figure}

We have solved the modified BD model with parameter values corresponding to disilicate
glass at $T=703$K ($k_{c}=18$) in Table I for three values $\mu = 10,$ 1, 1/3. The
corresponding modified free energy $e_{k}= g_{k}- \sum_{j=1}^{k-1}\Psi_{j}-\alpha$
is depicted in Fig.~\ref{figm1}. We observe that the energy barrier becomes flatter as $\mu$
decreases and the departure from the usual BDE with TF discrete diffusivity is greater. Fig.\
\ref{figm2}(a) shows the evolution of the size distribution function $s_{k}(t)$ towards
its stationary profile when $\mu=1$. Comparison between the three stationary profiles of the
size distribution function is depicted in Fig.~\ref{figm2}(b). Notice that the profiles for
$\mu=1$ and $\mu=1/3$ almost coincide and are noticeably less steep than the profile for
$\mu=10$.

\begin{figure}
\begin{center}
\includegraphics[width=8cm]{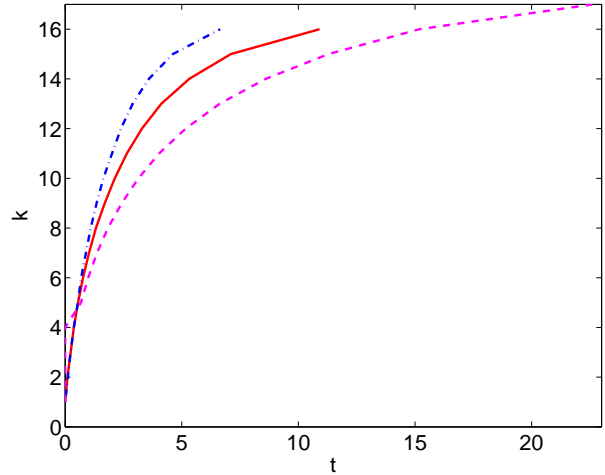}
\vspace{0.5 cm}
\caption{Evolution of the wave front position towards the critical size $k_{c}=18$ for
$\mu=10$ (dot-dashed line), $\mu=1$ (solid line) and $\mu=1/3$ (dashed line).}
\label{figm3}
\end{center}
\end{figure}

The size distribution function $s_{k}$ has a step-like profile whose inflection point marks
the instantaneous location of the wave front advancing in size space. Fig.~\ref{figm3}
shows the evolution of the wave front location. We observe that the velocity of the wave front
decreases as $\mu$ decreases.

\begin{figure}
\begin{center}
\includegraphics[width=8cm]{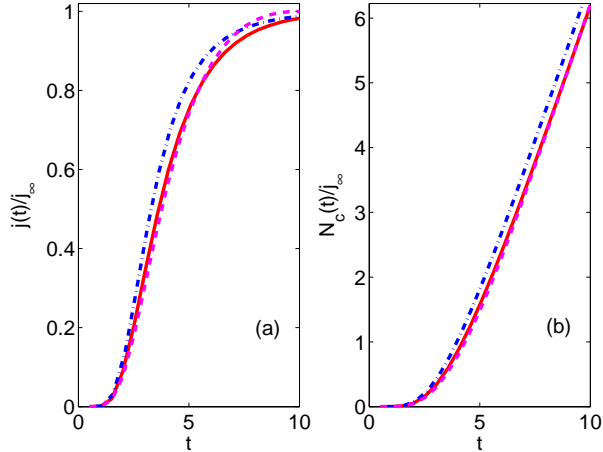}
\vspace{0.5 cm}
\caption{(a) Evolution of the normalized flux at critical size, $j(t)/j_{\infty}$, and (b)
number of clusters surpassing critical size, $N_{c}(t)$, for disilicate glass at 703 K and
$\mu=10$ (dot-dashed line), $\mu=1$ (solid line) and $\mu=1/3$ (dashed line). Time lags
are 2, 2.2 and 2.3, respectively. }
\label{figm4}
\end{center}
\end{figure}

Lastly, we have calculated the evolution of the flux at critical size and the number of
supercritical clusters for $\mu=10$, 1, 1/3; see Fig.~\ref{figm4}. For these values of $\mu$,
we have found $\varphi(0)= 4.79$, 6.9, 21.8 and $j_{\infty}= 2.1\times 10^{-9}$, $4.1
\times 10^{-6}$, $2.8\times 10^{-3}$, respectively.

\section{Discusion}
\label{sec:discusion}
In this paper, we have studied the case of phase segregation resulting when $\rho>\rho_{c}$.
Previously, other authors had carried out asymptotic studies of the BDE in the simpler
case of subcritical density, $\rho<\rho_{c}$, in which initial conditions of only monomers,
or more general ones, evolve towards the equilibrium distribution. In many cases of
polynomial growth for $d_{k}$, equilibrium is reached via a wave front profile for $s_{k}$,
which is similar to Eq.\ (\ref{a23}) with $A\propto K^\delta$, and $K\propto T^\mu$,
for appropriate positive $\delta$ and $\mu$; see Ref. ~\cite{kin02} and references cited
therein. This advancing and widening wave front leaves in its wake the equilibrium size
distribution.

In the more complex present case of phase segregation and indefinite
aggregate growth, a quasicontinuum wave front of $s_{k}$ emerges after a short transient 
which is governed by the discrete BDE. After this, the leading edge of the wave front
advances towards the critical size, and it slows down and stops
there, leaving behind it a quasi-equilibrium state. The arrival of
the wave front to the critical size marks the {\em ignition} of
nucleation of supercritical clusters, which ends when the stationary
Zeldovich rate is reached. Previous asymptotic theories (see 
\cite{tri87,shn87,shi90,dem93,mak00,shn01} and references cited therein) 
have been derived for the
continuum ZFE, not the discrete BDE, and thus their results
systematically misrepresent two things: (i) the time lags for
transient nucleation, as explained by Wu \cite{wu96}, and (ii) the
width of the wave front and the time to ignition in the nucleation
rate. The latter discrepancies occur because the diffusion
coefficient appearing in the continuum equation for the wave front
satisfies $D_{BDE}(K) = D_{ZFE}(K) - U(K)/2$, and therefore the time
to ignition in the nucleation rate for the BDE is {\em smaller} than
the corresponding one for the ZFE.

Let us briefly mention several existing asymptotic theories for the ZFE. Shneidman 
\cite{shn87} and Shi et al \cite{shi90} Laplace transformed 
the continuum ZFE and matched a first stage of pure advection of clusters to a local 
expansion about the wave front when it is near its final position at the critical size. They 
obtained our simplest formula for the nucleation rate, Eq.\ (\ref{a32}) with the same 
relaxation time, $\tau_{TF}$ or $\tau_{TDG}$, except that their values for $t_{M}$ were 
different from (\ref{a34}). This can be expected from Wu's arguments about approximating
the discrete BDE by the continuum ZFE \cite{wu96}; see the systematic shift of 
approximations of the ZFE with respect to numerical solutions of the BDE in  Fig.\ 20 of 
Ref.~\cite{wu96}. Trinkaus and Yoo \cite{tri87} studied a ZFE with a drift term 
linearized about the critical size (parabolic barrier) as an approximation to the full ZFE. 
Their results are comparable to those found by means of the Laplace transform and matched 
asymptotic expansions; see Wu's review \cite{wu96}. All these authors 
obtained a transition region for the nucleation rate $j(t)$ that was 
wider than observed in the numerical solution of the BDE. Several authors also found
a nucleation rate for supercritical clusters that did not tend to $j_{\infty}$ as $t\to\infty$
if $k\neq k_{c}$ \cite{tri87,shi90,dem93}, which is often called the {\em asymptotics 
catastrophe} \cite{mak00}. Our theory is free from this deficiency \cite{NBC}. 
Shneidman \cite{shn01} criticized Maksimov et al's 
result and extended his earlier asymptotic formula for the nucleation rate \cite{shn91} to 
non-critical sizes. The previous criticism of using approximations to the ZFE instead of 
approximations to the discrete BDE apply to these works.
 Our more precise approximation using Eq.\ (\ref{a25}) plus
the exact equations for the wave front location and its
instantaneous width improve upon other approximations and perform
better for materials with large critical sizes.

The time lag obtained from the numerical solution of the BDE with the TF diffusivity
(or from our asymptotic approximations using it) is too small as compared with experimental
results (about fifteen times smaller for disilicate at 703 K). The TF discrete diffusivity
yields an excessively small time unit, as shown in Table I. Another difficulty is that the TF
$d_{k}$ is proportional to the surface area of the $k$ cluster whereas diffusive growth of a
cluster after the nucleation has ceased produces $d_{k}\propto k^{1/3}$, as in late stage
coarsening theories \cite{LS}. We have proposed a new modification of the BDE that
yields a size advection velocity proportional to $k^{2/3}D_{+}g_{k}$ for small $k$, and to
$k^{1/3}D_{+}g_{k}$ for large $k$, thereby interpolating between TF and diffusive growth
formulas. With the new modification, time lags are somewhat larger, but $j_{\infty}$ 
increases unrealistically because the effective activation energy barrier diminishes. The 
usual way to improve agreement with experiments is to change the surface tension
so that $j_{\infty}$ agrees with measurements. It is clear that we should optimize the
choice of the surface tension and of the monomer-monomer bonding energy $\alpha$ to 
assess the merits of our proposal. Lastly, the flatness of the activation energy barrier for
small $\mu$ as depicted in Fig.~\ref{figm1} indicates that expansions about the barrier
maximum will provide poor approximations and that the corresponding asymptotic theory
for our modified BDE may not be an immediate extension of our results for the classical BDE.

\acknowledgments
The present work was financed by the Spanish MCyT grants BFM2002-04127-C02-01
and BFM2002-04127-C02-02, by the Spanish MECD grants MAT2005-05730-C02-01
and MAT2005-05730-C02-01, and by the European Union under grant
HPRN-CT-2002-00282.

\appendix
\section{Turnbull-Fisher discrete diffusivity}
\label{app:TF}
To derive $d_{k}$ for a glass-crystal transformation, we assume that the particles at the
glass-crystal interface are undergoing reactions
$$\mbox{glass} \rightleftarrows \mbox{crystal}.$$
The forward and backward reactions have activation energies $\varepsilon_{+}$ and
$\varepsilon_{-}$: During the conversion of one glass particle to crystal, one has to climb
up one side of an energy barrier with height $\varepsilon_{+}$, and then descend an
energy $\varepsilon_{-}$. The net free energy change should equal the chemical potential:
\begin{equation}
\frac{\varepsilon_{+}-\varepsilon_{-}}{k_{B}T} = D_{+}g_{k}. \label{tf1}
\end{equation}
The forward and backward rate constants of the reactions are
\begin{equation}
\kappa_{+}= \beta e^{-\frac{\varepsilon_{+}}{k_{B}T}}, \quad
\kappa_{-}= \beta e^{-\frac{\varepsilon_{-}}{k_{B}T}},   \label{tf2}
\end{equation}
respectively. Here $\beta$ is a positive constant. The kinetic rate constants can be rewritten
as
\begin{equation}
\kappa_{+}= \beta e^{-\frac{Q}{RT}} e^{-D_{+}g_{k}/2}, \quad
\kappa_{-}= \beta e^{-\frac{Q}{RT}}  e^{D_{+}g_{k}/2},   \label{tf3}
\end{equation}
provided $Q/N_{A} = (\varepsilon_{+}+\varepsilon_{-})/2$ is the mean barrier
height. The common prefactor $\beta e^{-Q/RT}$ is the inverse time
constant of the reactions when $D_{+}g_{k}=0$ and neither phase is
formed. For a spherical cluster with $k$ monomers, this prefactor is
the molecular jump rate at the cluster surface, which may coincide
with the rate of molecular diffusion in the glass phase. Since a
glass particle must push aside other glass particles when it moves,
it is plausible that molecular diffusion in the glass phase is an
activated process, as suggested by (\ref{tf3}). The diffusion
coefficient $D$ is one third the mean velocity times the mean free
path $\lambda\approx v^{1/3}$ according to elementary kinetic
theory. Then the diffusion jump rate is $3D/\lambda^2$, and we can
identify
\begin{eqnarray}
\beta e^{-\frac{Q}{RT}}= \frac{3D}{v^{2/3}},\quad\mbox{i.e.,}
 \quad
\beta = \frac{3D_{0}}{v^{2/3}}, \; && D=D_{0} e^{-\frac{Q}{RT}}.
\label{tf7}
\end{eqnarray}

Consider now the growth of a $k$-cluster by converting glass
particles to the crystal phase in the active sites at its interface.
Individual conversions are discrete events whose time sequence is a
Poisson process. Let $k_{+}$ be the number of glass to crystal
conversions happening on the cluster interface in a short time
$\delta t^*$. The expected value of $k_{+}$ is the number of active
sites at the cluster interface times the number of glass to crystal
conversions $\kappa_{+}\delta t^*$. The number of active sites is
the area of the spherical $k$-cluster of radius $[3kv/(4\pi)]^{1/3}$
divided by the area a single molecule would occupy on the cluster
interface, namely, $\pi[3v/(4\pi)]^{2/3}$. Then, ${\cal A}_{k} = 4
k^{2/3}$ and $\langle k_{+}\rangle$ satisfies
\begin{eqnarray}
\langle k_{+}\rangle = {\cal A}_{k} \kappa_{+} \delta t^*, \;
\langle (k_{+}-\langle k_{+}\rangle)^2\rangle = \langle k_{+}\rangle
= {\cal A}_{k} \kappa_{+} \delta t^*. \label{tf10}
\end{eqnarray}
The conversions from crystal to glass are another Poisson process for which
\begin{eqnarray}
\langle k_{-}\rangle = {\cal A}_{k} \kappa_{-} \delta t^*, \;
\langle (k_{-}-\langle k_{-}\rangle)^2\rangle = \langle k_{-}\rangle
= {\cal A}_{k} \kappa_{-} \delta t^*. \label{tf12}
\end{eqnarray}
We assume that the backward and forward Poisson processes are independent, so
\begin{eqnarray}
\langle (k_{+}-\langle k_{+}\rangle)(k_{-}-\langle k_{-}\rangle)\rangle = 0.
\label{tf13}
\end{eqnarray}
Equations (\ref{tf10}) to (\ref{tf13}) imply that the net number of
glass-crystal conversions, $\delta k\equiv k_{+}-k_{-}$, has the
statistics
\begin{eqnarray}
\langle \delta k\rangle = {\cal A}_{k} (\kappa_{+}-\kappa_{-})\,
\delta t^*,  \;\; \langle (\delta k-\langle \delta
k\rangle)^2\rangle = {\cal A}_{k}
 (\kappa_{+}+\kappa_{-})\,  \delta t^*. \label{tf15}
\end{eqnarray}
Inserting (\ref{tf3}) and (\ref{tf7}) into these equations, we
obtain
\begin{eqnarray}
\langle \delta k\rangle = \frac{24 k^{2/3} D}{v^{2/3}}
\sinh\left(\frac{- D_{+}g_{k}}{2}\right)\, \delta t^*,    \;\;
\langle (\delta k-\langle \delta k\rangle)^2\rangle = \frac{24
k^{2/3} D}{ v^{2/3}} \cosh\left(\frac{D_{+}g_{k}}{2}\right)\, \delta
t^*. \label{tf16}
\end{eqnarray}
We now identify the velocity $v_{k}= (e^{-D_{+}g_{k}}-1)\, d_{k}$ in the BDE with
$\Omega^{-1}\delta\langle k\rangle/\delta t^*$, and this yields the diffusivity
(\ref{e18}). Eq.\ (\ref{tf16}) also provides the diffusion coefficient (\ref{a17}).

\end{document}